\documentclass[%
 preprint,
 amsmath,amssymb,
 aps,
]{revtex4-1}

\usepackage[pdftex]{graphicx}
\usepackage{rotating}
\usepackage{multirow}

\usepackage{amsmath}
\usepackage{amssymb}
\usepackage{psfrag}
\usepackage{multirow}
\usepackage{rotating}
\usepackage{array}
\usepackage{units}

\begin{document}

\title{Nucleated polymerisation in the presence of pre-formed seed filaments}

\author{Samuel~I.~A.~Cohen$^1$, Michele Vendruscolo$^1$, 
Christopher~M.~Dobson$^1$, and Tuomas P.~J.~Knowles$^{1}$}

\affiliation{$^1$Department of Chemistry, University of Cambridge, Lensfield
Road, Cambridge CB2 1EW, UK}

\date{\today}

\begin{abstract}We revisit the classical problem of nucleated polymerisation and 
derive a range of exact results describing polymerisation in systems 
intermediate between the well-known limiting cases of a reaction starting from 
purely soluble material and for a reaction where no new growth nuclei are 
formed.\end{abstract}

\maketitle

\section{Introduction}

The classical theory of nucleated polymerisation\cite{Oosawa1975} describes the 
growth of filamentous structures formed through homogeneous 
nucleation\cite{Gibbs1878, Volmer1926, Kaishev1934, Stranski1935, Becker1935, 
Avrami1939}. This framework was initially developed by Oosawa and coworkers in 
the 1960s\cite{Oosawa1975, Oosawa1962} to describe the formation of 
biofilaments, including actin and tubulin. This theory has been generalised to 
include secondary nucleation processes by Eaton and Ferrone\cite{Ferrone1985b} 
in the context of their pioneering work elucidating the polymerisation of sickle 
haemoglobin, and by Wegner\cite{Wegner1975} in order to include fragmentation 
processes into the growth model for actin filaments. 

For irreversible growth in the absence of pre-formed seed material and secondary 
nucleation pathways, in 1962 Oosawa presented solutions to the kinetic equations 
which were very successful in describing a variety of characteristics of the 
polymerisation of actin and tubulin. The other limiting case, namely where seed 
material is added at the beginning of the reaction and where no new growth 
nuclei are formed during the reaction, is also well known. In this paper, we 
present exact results which encompass all cases between these limiting 
scenarios, extending the results of Oosawa for a system dominated by primary 
nucleation to the case where an arbitrary concentration of pre-formed seed 
material is present.  We also discuss a range of general closed form results 
from the Oosawa theory for the behaviour of a system of biofilaments growing 
through primary nucleation and elongation. We then compare the behaviour of 
systems dominated by primary nucleation to results derived recently for systems 
dominated by secondary nucleation.


\section{Results and Discussion}
\subsection{Derivation of the rate laws for the polymer number and mass 
concentrations}

The theoretical description of the polymerisation of proteins such as actin and 
tubulin to yield functional biostructures was considered in the 1960s by 
Oosawa\cite{Oosawa1962}.  For a system that evolves through primary nucleation 
of new filaments, elongation of existing filaments, and depolymerisation from 
the filament ends, the change in concentration of filaments of size $j$, denoted 
$f(j,t)$, is given by the master equation\cite{Oosawa1962, Oosawa1975}:
\begin{equation} \begin{split} \frac{\partial 
f(t, j)}{\partial t} = &\; 2 m(t) k_+ f(t, j-1)- 2 m(t) k_+ f(t, j)\\
 & +2 k_\mathrm{off} f(t,j+1)-2 k_\mathrm{off} f(t,j)\\
&+ k_n m(t)^{n_c} \delta_{j,n_c}
\label{eq:masteroosawa}
\end{split}
\end{equation}
where $k_+$, $k_\mathrm{off}$, $k_n$ are rate constants describing the 
elongation, depolymerisation and nucleation steps and $m(t)$ is the 
concentration of free monomeric protein in solution. The factor of 2 in 
Eq.~\eqref{eq:masteroosawa} originates from the assumption of growth from both 
ends. For the case of irreversible biofilament growth, the polymerisation rate 
dominates over the depolymerisation rate; from Eq.~\eqref{eq:masteroosawa}, the 
rate of change of the number of filaments, $P(t)$, and the free monomer 
concentration, $m(t)$, were shown by Oosawa under these conditions 
\cite{Oosawa1962, Oosawa1975} to obey:
\begin{align}
\frac{dP}{dt} &= k_n m(t)^{n_c}
\label{eq:oosawaP}\\
\frac{dm}{dt} &= -2k_+ m(t) P(t)
\label{eq:oosawam}
\end{align}
Combining Eqs.~\eqref{eq:oosawaP} and \eqref{eq:oosawam} yields a differential 
equation for the free monomer concentration\cite{Oosawa1975}:
\begin{equation}
-\frac{d^2}{dt^2}\mathrm{log}(m(t))=2k_+k_n m(t)^{n_c}
\end{equation}. 

Here, we integrate these equations in the general case where the initial state 
of the system can consist of any proportion of monomeric and fibrillar material; 
this calculation generalises the results presented by Oosawa to include a finite 
concentration of seed material present at the start of the reaction.  Beginning 
with Eqs.~\eqref{eq:oosawaP} and \eqref{eq:oosawam}, the substitution $z(t) := 
\mathrm{log}(m(t))$ followed by multiplication through by $dz/dt$ yields:
\begin{equation}
-\frac{d}{dt}\left[\frac{n_c}{4k_+k_n} \left(\frac{dz}{dt}\right)^2\right] = 
\frac{d}{dt}e^{n_c z}
\end{equation}
Integrating both sides results in:
\begin{equation}
-\frac{n}{2} \left(\frac{dz}{dt}\right)^2 = 2k_+k_n e^{n_c z} + A = -\frac{d^2 
z}{dt^2} + A
\end{equation}
we obtain a separable equation for $dz/dt$, which can be solved to yield:
\begin{equation}
\frac{dz}{dt}=\sqrt{\frac{2 A}{n_c}}\mathrm{tanh}\left(\frac{\sqrt{2A n_c}}{2} 
(-t+2 B)\right)
\end{equation}
Integration and exponentiation yields the expression for $m(t)$:
\begin{equation}
m(t)=\left[\frac{A}{2 k_+ k_n}\mathrm{sech}\left(\sqrt{\frac{A 
n_c}{2}}(t-2B)\right)^2\right]^{1/n_c}
\end{equation}
Inserting the appropriate boundary conditions in terms of $m(0)$ and $P(0)$ 
fixes the values of the constants $A$ and $B$, resulting in the final exact 
result for the polymer mass concentration $M(t) = m_\mathrm{tot} - m(t)$:
\begin{equation}
M(t)= m_{\mathrm{tot}}-m(0)\left[\mu\, \mathrm{sech}\left(\nu+ \lambda _0 \beta^{-\frac{1}{2}}\mu  
t\right)\right]^{\beta}
\label{eq:oosawaseededM}
\end{equation}
where the effective rate constant $\lambda$ is given by $\lambda=\sqrt{2
k_n k_+ m(0)^{n_c}}$ and $\beta =2/n_c$, $\mu =\sqrt{1+\gamma^2}$, 
$\nu = \text{arsinh}\left(\gamma\right)$ for $\gamma =2k_+ P(0)/(\beta^{\frac{1}{2}}\lambda)$.  

We note that this expression only depends on two combinations of the microscopic 
rate constants, $k_0 = 2k_+ P(0)$ and $\lambda$. The result reveals that 
$\lambda$ controls the aggregation resulting from the newly formed aggregates, 
whereas $k_0$ defines growth from the pre-formed seed structures initially 
present in solution. In the special case of the aggregation reaction starting 
with purely soluble proteins, $P(0) = 0$, $m(0) = m_\mathrm{tot}$, these 
expressions reduce to $\mu \rightarrow 1$ and $\nu \rightarrow 0$, and 
Eq.~\eqref{eq:oosawaseededM} yields the result presented by 
Oosawa\cite{Oosawa1975} and the single relevant parameter in the rate equations 
is $\lambda$. Interestingly, generalisations of Eq.~\eqref{eq:oosawaseededM} 
which include secondary pathways, maintain the dependence on $\lambda$ and $k_0$ 
but introduce an additional parameter analogous to $\lambda$ for each active 
secondary pathway\cite{Cohen2011_1, Cohen2011_2, Cohen2011_3, Knowles2009}.

An expression for the evolution of the polymer number 
concentration, $P(t)$ may be derived using Eq.~\eqref{eq:oosawaseededM}.  Direct 
integration of Eq.~\eqref{eq:oosawaP} gives the result for $P(t)$:
\begin{equation}
P(t) = P(0)+k_n m(0)^{n_c} \mu\frac{\mathrm{tanh}(\nu+\beta^{-\frac{1}{2}}\lambda \mu t)-\mathrm{tanh}(\nu)}{\beta^{-\frac{1}{2}}\lambda }
\label{eq:oosawaseededP}
\end{equation}
Eqs.~\eqref{eq:oosawaseededP} and \eqref{eq:oosawaseededM} give in closed form the time evolution of the biofilament number and mass concentration growing through primary nucleation and filament elongation.

\begin{figure}[tb]
\begin{center}
\includegraphics[width=1.0\textwidth]{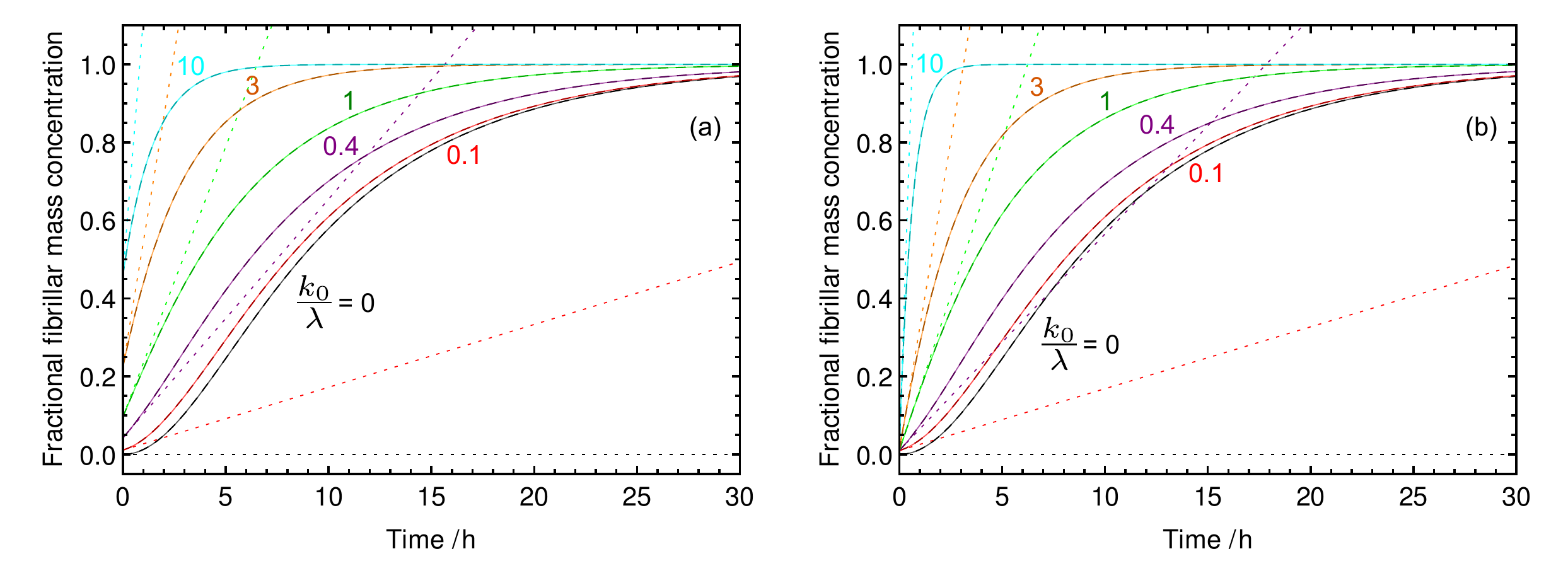}
\caption{Nucleated polymerisation in the presence of seed material.  The thick dashed lines are the exact solution to the rate equations Eq.~\eqref{eq:oosawaseededM}; the thin solid lines are calculated from numerical simulations of the master equation Eq.~\eqref{eq:masteroosawa}.  The dotted lines are the initial gradients $dM/dt|_{t=0} = M(0) + 2k_+m(0)P(0)t$; a lag-phase exists when the initial gradient is not the maximal gradient.  The numbers accompanying each curve are $k_0/\lambda$; Eq.~\eqref{eq:k0} predicts that a lag-phase only exists when this ratio is less than unity.  (a): Polymerisation in the presence of an increasing quantity of 
seed material of a fixed average length (5000 monomers per seed)  added at the beginning of the reaction.  The
seed concentrations given as a fraction of the total concentration of monomer present are right to left):
0, 0.01, 0.04, 0.1, 0.2, 0.5. (b): Nucleated polymerisation in the presence of a fixed quantity (1\% of total monomer in the system) of 
seed material of varying average length.  The
average number of monomer per seed are (right to left):
N/A (unseeded), 5000, 1000, 500, 200, 50.  The other parameters for both panels are: $m_\mathrm{tot} = 10\mu M$, $n_c = 3$, $k_n m_\mathrm{tot}^{n_c -1} = 1\cdot10^{-9}s^{-1}$, $k_+ = 1\cdot10^5$M$^{-1}$ s$^{-1}$.
\label{fig:seeded}}
\end{center}
\end{figure}

\subsection{Characteristic features of growth involving pre-formed seed 
material}

Insight into the early time behaviour of the polymer mass concentration can be 
obtained by expanding Eq.~\eqref{eq:oosawaseededM} for early times to yield:
\begin{equation}
M(t) \xrightarrow{t \rightarrow 0} M(0) + k_0 m(0) t + m(0)[\lambda^2-k_0^2]t^2 
/2 +\mathcal{O}(t^3)
\end{equation}
This expression recovers the characteristic $\sim t^2$ dependence of the Oosawa 
theory and has an additional term linear in time relating to the growth of 
pre-formed aggregates.

In many cases, Eq.~\eqref{eq:oosawaseededM} describes a sigmoidal function with 
a lag phase.  The time of maximal growth rate, $t_\mathrm{max}$, can be found 
from the inflection point of the sigmoid from the condition $d^2M/dt^2 = 0$:
\begin{equation}
t_\mathrm{max} = 
\left[\mathrm{artanh}\left(\sqrt{\frac{1}{1+\beta}}\right)-\mathrm{arsinh}\left(\gamma\right)\right]( \mu \beta^{-\frac{1}{2}}\lambda_0)^{-1}
\end{equation}
such that a lag phase exists only for:
\begin{equation}
\mathrm{artanh}\left(\sqrt{\frac{1}{1+\beta}}\right) > 
\mathrm{arsinh}\left(\gamma\right)
\end{equation}
Using the composition $\mathrm{sinh}(\mathrm{artanh}(x)) = x/\sqrt{1-x^2}$ reduces this to the simple condition:
\begin{equation}
k_0 < \lambda
\label{eq:k0}
\end{equation}
In other words, a point of inflection exists if the growth through elongation 
from the ends of pre-existing seeds, $k_0$, is less effective that the effective 
growth through nucleation and elongation of new material, $\lambda$.  This 
result imples that an increased nucleation rate promotes the existence of an 
inflection point, whereas an increased elongation rate or an increased level of 
seeding tends to disfavour its existence.  In particular, we also note that in 
the absence of nucleation, an inflection point cannot exist in the polymer mass 
concentration as a function of time. Interestingly, the result Eq.~\eqref{eq:k0} 
is analogous to the criterion applicable for fragmentation dominated growth 
where a lag phase only exists when the parameters controlling 
fragmentation-related secondary nucleation is larger than $e k_0$.

The maximal growth rate, $r_\mathrm{max}$, is given by:
\begin{equation}
r_\mathrm{max} = \frac{2 m(0)}{\sqrt{n_c(2+n_c)}} \left(\frac{2 \mu^2}{2+n_c}\right)^{\frac{1}{n_c}} \mu \lambda
\label{eq:r}
\end{equation}
which occurs at a polymer mass concentration $M_\mathrm{max}$ given from Eq.:
\begin{equation}
M(t_\mathrm{max}) = m_\mathrm{tot} - m(0) \mu^{\frac{2}{n_c}} \left(1-\frac{n_c^2}{(2+n_c)^2}\right)^\frac{1}{n_c}
\end{equation}
The lag time, $\tau_\mathrm{lag} := t_\mathrm{max} - M(t_\mathrm{max})/r_\mathrm{max}$, is then given by:
\begin{equation}
\begin{split}
\tau_\mathrm{lag} =& 
\left[\mathrm{artanh}\left(\sqrt{\frac{1}{1 + \beta}}\right)-\mathrm{arsinh}\left(\gamma\right) \right. \\&\left.- \frac{m_\mathrm{tot} - m(0) 
\mu^{\frac{2}{n_c}} 
\left(1-\frac{n_c^2}{(2+n_c)^2}\right)^\frac{1}{n_c}}{\frac{2 
m(0)}{\sqrt{n_c(2+n_c)}} \left(\frac{2 
\mu^2}{2+n_c}\right)^{\frac{1}{n_c}}}\right]( \mu \beta^{-\frac{1}{2}} \lambda)^{-1}
\end{split}
\label{eq:oo2}
\end{equation}
Interestingly, from Eq.~\eqref{eq:oo2}, we note that a point of inflection can 
never exist for $P(t)$ for simple nucleated polymerisation. By contrast, when 
secondary pathways are active, an inflection point can frequently be 
present\cite{Cohen2011_2}.

\begin{table}[tb]
 \begin{tabular}{>{\centering}p{5.0cm}|>{\centering}p{3.0cm}|>{\centering}p{3.0cm}|>{\centering}p{4.7cm}}
\hline
&\textbf{Primary  
nucleation}&\textbf{Fragmentation}&\textbf{Monomer-dependent secondary nucleation}\tabularnewline
\hline
\textbf{Kinetic parameters}&$\lambda$, $k_+$&$\lambda$, $\kappa_-$, $k_+$&$\lambda$, $\kappa_2$, $k_+$\tabularnewline
\hline
\textbf{Early time growth}&Polynomial&Exponential&Exponential\tabularnewline
\hline
\textbf{Scaling behaviour\\ (lag time, max growth rate)}&Yes\\with $\lambda$&Yes\\with $\kappa_-$&Yes\\with $\kappa_2$ \tabularnewline
\hline
\textbf{Positive feedback}&No&Yes&Yes
 \tabularnewline
\hline
\end{tabular}
 \caption{Comparison of biofilament growth dominated by primary and secondary nucleation pathways}
\label{tab:exponents}
\end{table}

\subsection{Comparison between nucleated polymerisation in the presence and 
absence of secondary pathways}

Many systems that evolve through nucleated polymerisation display characteristic 
scaling behaviour\cite{Cohen2011_1, Cohen2011_2, Cohen2011_3, Knowles2009, 
Faendrich2007}. This behaviour can be seen to be a consequence of the fact that 
under many conditions, the rate equations are dominated by a single parameters 
that corresponds to the dominant form of nucleation: $\lambda$ for classical 
nucleated polymerisation and $\kappa_{2,-}$ for polymerisation in the presence 
of secondary pathways. These parameters have the general form $\sqrt{2 k_+ m(0) 
k_N m(0)^n}$ where $k_N = {k_n, k_-, k_2}$ corresponds to the nucleation process 
and $n$ is related to the monomer dependence of this process: $n=n_c-1$, where $n_c$ is the 
critical nucleus size for primary nucleation, $n=0$ for fragmentation driven 
growth and $n=n_2$, the secondary nucleus size in cases where monomer-dependent 
secondary nucleation is dominant. The dominance of a single combination of the 
rate constants implies that many of the macroscopic system observables will be 
correlated since they are dependent on the same parameter. A striking examples 
of this behaviour is provided by the very general correlation between the 
lag-time and the maximal growth rate\cite{Knowles2009, Faendrich2007}, which is 
manifested in the present case in Eqs.~\eqref{eq:r} and \eqref{eq:oo2} as $r_\mathrm{max} \sim \lambda$ and $\tau_\mathrm{lag} \sim \lambda^{-1}$.

Interestingly the rate equations describe sigmoidal curves both in the presence 
and in the absence of secondary nucleation processes. For more complex primary 
nucleation pathways\cite{Flyvbjerg1996, Ferrone1999} the polynomial form for the 
early time solution is maintained, but higher-order exponents are obtained. In 
the absence of secondary processes, however, the lag-phase is less marked since 
the early time rise is a slower polynomial relationship rather than the 
exponential onset characteristic of secondary pathways\cite{Ferrone1999}. This 
observation implies that the difference between a high-order polynomial and an 
exponential may not be apparent in experimental data in the presence of noise, 
and therefore a global analysis of the system under different conditions is 
required in order to obtain robust mechanistic information\cite{Knowles2009}.

\section{Conclusion}
In this paper, we have provided results for the time course of nucleated 
polymerisation for systems that are initially in a mixed state and contain both 
monomeric and fibrillar material. These results generalise the classical Oosawa 
theory that describes the formation of biofilaments to cases where an arbitrary 
amount of pre-formed seed material is present in the system. Furthermore, these 
results represent a reference to which polymerisation driven by secondary 
pathways can be compared. 

\section*{Acknowledgements}

We are grateful to the Schiff Foundation (SIAC), and to the Wellcome (MV, CMD, 
TPJK) and Leverhulme Trusts (CMD) for financial support.

\bibliographystyle{mdpi}
\makeatletter
\renewcommand\@biblabel[1]{#1. }
\makeatother


\end{document}